\documentclass[sigconf,screen]{acmart}

\usepackage[frozencache,cachedir=.]{minted}
\usepackage{listings}
\usepackage{multirow}

\AtBeginDocument{%
  }

\setcopyright{rightsretained}
\copyrightyear{2025}
\acmYear{2025}
\acmConference[]{Under Review}{2025}{Anonymous Location}
\acmBooktitle{Under Review}

\begin{document}

\title{Refactoring Detection in C++ Programs with RefactoringMiner++}

\author{Benjamin Ritz}
\authornote{These authors contributed equally to this work.}
\orcid{0009-0000-7774-6693}
\email{benjamin.ritz@tugraz.at}
\affiliation{%
  \institution{Graz University of Technology}
  \city{Graz}
  \country{Austria}
}

\author{Aleksandar Karaka\v{s}}
\authornotemark[1]
\orcid{0000-0002-6195-4962}
\email{a.karakas@fh-kaernten.at}
\affiliation{%
  \institution{Carinthia University of Applied Sciences}
  \city{Villach}
  \country{Austria}
}

\author{Denis Helic}
\orcid{0000-0003-0725-7450}
\email{dhelic@tugraz.at}
\affiliation{%
  \institution{Graz University of Technology}
  \city{Graz}
  \country{Austria}
}

\renewcommand{\shortauthors}{Ritz et al.}

\begin{abstract}
  Commits often involve refactorings---behavior-preserving code modifications aiming at software design improvements. Refactoring operations pose a challenge to code reviewers, as distinguishing them from behavior-altering changes is often not a trivial task. Accordingly, research on automated refactoring detection tools has flourished over the past two decades, however, the majority of suggested tools is limited to Java projects. In this work, we present RefactoringMiner++, a refactoring detection tool based on the current state of the art: RefactoringMiner 3. While the latter focuses exclusively on Java, our tool is---to the best of our knowledge---the first publicly available refactoring detection tool for C++ projects. RefactoringMiner's thorough evaluation provides confidence in our tool's performance. In addition, we test RefactoringMiner++ on a small seeded dataset and demonstrate the tool's capability in a short demo involving both refactorings and behavior-altering changes. A screencast demonstrating our tool can be found at\\\url{https://cloud.tugraz.at/index.php/s/oCzmjfFSaBxNZoe}.
\end{abstract}

\ccsdesc[500]{Software and its engineering~Software maintenance tools}

\keywords{Refactoring Detection, C++, RefactoringMiner}


\maketitle

\section{Introduction}

Refactoring is the process of improving the design of existing code without changing its behavior \cite{fowler2018refactoringCatalogue}. Developers perform refactorings frequently \cite{tose11howWeRefactor}, thus attaining benefits, such as reduced code duplication as well as improved readability, maintainability and testability \cite{tose14refactoringStudyAtMicrosoft,fse16whyWeRefactor}.
Commits which include both refactorings and behavior-altering code modifications---so-called \emph{tangled commits} \cite{msr13untanglingCodeChanges,saner15untanglingCodeChanges}---are a widespread phenomenon \cite{tose11howWeRefactor,ecoop12studyCodeEvolutionWithCodingTracker}. Such commits are problematic, as they make code review and integration harder \cite{fse12understandCodeChanges,thesis12integrationOfCommits}. 
Tangled commits also entail repercussions for analyses of commit histories \cite{msr13untanglingCodeChanges}. Untangling commits requires identifying refactorings, which is a difficult task given that developers tend to leave refactoring operations undocumented \cite{tose11howWeRefactor}. 
This problem calls for tools which are capable of detecting refactorings.
Such tools also help generate new insights into developers' motivation for refactorings. This, in turn, can lead to innovations in the field of refactoring recommendation systems \cite{fse16whyWeRefactor}.

Moreover, refactoring detection also has a major impact on CS education where many programming courses feature group projects \cite{sigcse20assessingIndividualContributionsWithGitLogs,cse20githubCaseStudy}. These courses pose a challenge in terms of grading \cite{sigcse20assessingIndividualContributionsWithGitLogs} and, thus, several tools exist to help instructors measure individual contributions in a student team \cite{gitreporter}. However, these tools are based on the number of contributed lines of code. A tool which further analyzes the nature of these contributions by distinguishing refactorings from other changes could support instructors in assessing students' work. Especially instructors in courses with an emphasis on refactoring such as the maintenance courses (cf. Petrenko et al.~\cite{computer07teachingSoftwareEvolution}) might benefit from automated refactoring detection.

In the past two decades, several tools emerged for detecting refactorings. The majority of efforts resulted in new approaches to analyze Java projects (e.g., \cite{iwpse04javaRefactoringDetectionWithCosineSimilarity,ecoop06javaRefactoringCrawler,ase06javaRefactoringDetection,ieeeIcsm10javaTemplateBasedReconstruction,fse10ref-finder,ieeeMsr17refdiff,ieeeTose20refdiff2multilang,icse18refactoringMiner,tse20refactoringMiner2,tosem24refactoringMiner3,ieeeAccess21refdetect}). Due to the need for finding refactorings in other programming languages, recent years also saw research on refactoring detection for software written in Python \cite{pythonAdaptedRefactoringMiner,pyref,bscThesis24pythonWithRefDetect}, JavaScript \cite{ieeeTose20refdiff2multilang}, Go \cite{RefDiff4Go}, Kotlin \cite{ase21javaKotlinRefactoringInJetBrainsIde,bscThesis23kotlinWithRefDetect,saner24kotlinWithRefDetect}, C \cite{ieeeTose20refdiff2multilang}, and C++ \cite{ieeeAccess21refdetect}. Some of these tools \cite{bscThesis24pythonWithRefDetect,bscThesis23kotlinWithRefDetect,saner24kotlinWithRefDetect}---including the one for C++ projects---are based on RefDetect \cite{ieeeAccess21refdetect}, which ``is not currently publicly available'' \cite{RefDetectWebsite}.

In this work, we present RefactoringMiner++, an open-source refactoring detection tool based on the current state of the art: RefactoringMiner 3 \cite{tosem24refactoringMiner3}. While the latter focuses exclusively on Java, our tool is---to the best of our knowledge---the first publicly available refactoring detection tool for C++ projects.\footnote{Source code, illustrative example, evaluation dataset, and evaluation results are available on GitHub at \url{https://github.com/benzoinoo/RefactoringMinerPP}.}


\section{Related Work}

\noindent\textbf{Tools for Java Projects.} We first focus on two state-of-the-art tools for refactoring detection in Java projects: RefDetect~\cite{ieeeAccess21refdetect} and RefactoringMiner 3~\cite{tosem24refactoringMiner3}. RefDetect's authors compared their tool to RefactoringMiner 2 \cite{tse20refactoringMiner2} and report a slightly better performance. In RefDetect, each class is summarized using a string. Strings representing successive commits are aligned using the FOGSAA algorithm~\cite{nature13stringAlignmentAlgorithm}. The aligned strings are then used to match parts of two commits. RefDetect can also find refactorings in C++ projects, however, the tool is not publicly available \cite{RefDetectWebsite}. 

RefDetect's number of parameters can also be considered as a drawback \cite{tse20refactoringMiner2}. The FOGSAA algorithm has three parameters, while the rest of the refactoring detection algorithm requires setting six threshold values for tweaking detection sensitivity. The RefDetect authors offer defaults for these settings, however, ``it is very difficult to derive \emph{universal} threshold values that can work well for all projects'' \cite{tse20refactoringMiner2}.
Calibration also requires an extensive dataset of refactorings. While such data exist for Java projects (e.g., \cite{icse18refactoringMiner,tse20refactoringMiner2}), it does not yet seem to be the case for C++ \cite{ieeeAccess21refdetect}. 
As our tool is based on RefactoringMiner, it does not require such settings \cite{tosem24refactoringMiner3}.

Furthermore, RefDetect is restricted to 27 object-oriented refactoring types, but does not cover low-level refactorings such as renaming variables \cite{ieeeAccess21refdetect}.
RefactoringMiner 3.0, on the other hand, can detect more than 100 refactoring types---even some which are unrelated to classes \cite{tosem24refactoringMiner3}. While a part of the supported refactoring types are Java-specific, we expect many to carry over to C++.

As RefDetect, RefactoringMiner 3 was compared to version 2 of RefactoringMiner. The new version outperforms its predecessor in terms of matching parts of two commits. Large performance differences were observed in cases with one-to-many mappings and in connection with refactorings. Reasons for the increased performance are the addition of many refactoring types, an enhanced matching algorithm, and the new abstract syntax tree differencing (AST diff) feature. AST diff allows RefactoringMiner 3 to perform more fine-grained code comparisons. Being refactoring-aware and language-specific, the AST diff feature of RefactoringMiner 3 outperforms even state-of-the-art AST diff tools \cite{tosem24refactoringMiner3} such as GumTree \cite{tose32gumtree3,icse24gumtree3}. RefactoringMiner's authors propose their tool's extension to other programming languages as future research topic \cite{tosem24refactoringMiner3}. Our work addresses this research gap. We chose C++, which shares many characteristics with Java. For instance, both are statically and strongly typed and offer object-orientation with inheritance.

\noindent\textbf{Other languages.} By replacing the language-dependent part of RefDetect, it can be adapted to new programming languages. C++ was chosen by the RefDetect developers as the second language and they report an F\textsubscript{1} score of $0.95$ on an unpublished dataset with refactoring commits created 
by students \cite{ieeeAccess21refdetect}. RefDetect's authors acknowledge that the dataset used for evaluation is not comprehensive and limited to object-oriented refactoring operations. C++, however, also allows for other programming styles, such as structured programming or template metaprogramming \cite{meyers05effectiveCpp}.

While promising results with an F\textsubscript{1} score of $0.84$ were also reported for an extension of RefDetect to the Kotlin language, its authors also highlight differences between Kotlin and Java, RefDetect's primary language \cite{saner24kotlinWithRefDetect}. Language incompatibilities also affect the RefDetect extension for Python and, thus, this tool lacks support for many Python features, such as list comprehensions, module imports, decorators, properties, and multiple inheritance \cite{bscThesis24pythonWithRefDetect}.

There has also been work on extending RefactoringMiner 2 to other languages, such as Python ~\cite{pythonAdaptedRefactoringMiner} and Kotlin \cite{ase21javaKotlinRefactoringInJetBrainsIde}. 
Python-adapted RefactoringMiner uses Jython to bring Python programs to the Java Virtual Machine, where they can be analyzed by RefactoringMiner. The reported 29 \cite{pythonAdaptedRefactoringMiner} 
and 19 \cite{ase21javaKotlinRefactoringInJetBrainsIde} supported refactoring types do not reach the 40 refactoring types RefactoringMiner 2 can detect.

Another refactoring detector for Python is PyRef \cite{pyref}. Instead of using RefactoringMiner, it implements RefactoringMiner's algorithm in Python. PyRef outperforms Python-adapted RefactoringMiner, but is restricted to nine method-related refactoring types.

In our approach, a modification of RefactoringMiner 3 is used.

\noindent\textbf{Datasets for Tool Evaluation.} While some works on refactoring detection tools don't include an evaluation of the tool's accuracy at all (e.g., \cite{ase21javaKotlinRefactoringInJetBrainsIde}), other authors evaluate their tool on a dataset comprising commits with a ground truth of applied refactorings for each commit. Such a dataset is sometimes generated by collecting refactoring operations performed by students (e.g., \cite{ieeeMsr17refdiff,ieeeAccess21refdetect}). An advantage of this approach is that the ground truth is known. However, these \emph{seeded} refactorings do not necessarily reflect real-world development practice and may be too easy to detect \cite{tse20refactoringMiner2}.

A different approach to constructing a refactoring dataset is to mine commits of open-source projects and identify the refactorings contained in these commits. This process is labor-intensive, as it requires experts to determine the ground truth of applied refactorings. 
A common limitation incurred when evaluating tools on such datasets is the lack of independent experts who shape the ground truth, thus, potentially causing experimenter bias \cite{tse20refactoringMiner2,ieeeAccess21refdetect}. Nonetheless, such datasets are invaluable---due to their size and because they reflect the refactoring practices in the collected projects.

Usually, the first step to produce such a dataset involves running multiple refactoring detectors on a set of commits (e.g., \cite{pyref,saner24kotlinWithRefDetect,icse18refactoringMiner,tse20refactoringMiner2}). Refactorings reported by any of the used tools form the list of candidates. This list is then scrutinized by experts to determine which candidates are actual refactorings. Due to the lack of other publicly available C++ refactoring detectors, this approach to build an evaluation dataset is infeasible. Instead, we compare our tool to its Java counterpart. For this comparison, we generate a small seeded dataset of equivalent refactorings in C++ and Java using a large language model (LLM).

\section{Design and Implementation}

RefactoringMiner 3 uses the Eclipse JDT parser to analyze programs written in Java. Given Java source code as input, the AST parser generates an abstract syntax tree, which is a hierarchical representation of the program. By traversing this AST, RefactoringMiner constructs an in-memory model representation of the Java program. This model contains all key elements of the program, including classes, attributes, methods, statements, or expressions. During refactoring detection, two commits in a Java project are compared. For each commit, a model is constructed and refactoring detection is based on these models rather than the source code directly. Hence, RefactoringMiner's algorithms are---to some degree---language-independent. If a model of a program can be constructed, RefactoringMiner should be able to analyze it, regardless of the programming language. However, the model used by RefactoringMiner is based on Java. This language-specificity contributes to RefactoringMiner's state-of-the-art performance \cite{tosem24refactoringMiner3}, but requires a certain degree of similarity between Java and the desired programming language. C++ fulfills this requirement.

We built a tool written in C++ that takes the source code of a C++ program as input and constructs the corresponding model for RefactoringMiner. To create a model, we traverse the program’s AST using Clang. Clang is a widely used and actively maintained open-source compiler for C, C++ and Objective-C. To obtain access to the parsed AST, we use libClang, the ``most mature and stable'' \cite{sede14libClang} interface to Clang. 
Once the models are constructed for both commits of interest, we move them from our C++ program to RefractoringMiner.
To this end, we serialize the models using the human-readable JSON format. We adapted RefactoringMiner to skip the usual model construction phase for Java programs and to directly use the JSON models instead. Using these models, RefatoringMiner is unaware that it is analyzing C++ programs. Hence, our modified version of the RefactoringMiner is able to detect many of the refactoring types originally supported by RefactoringMiner.

Furthermore, we extend RefactoringMiner to report behavior-altering changes in addition to refactorings.

\section{Illustrative Example} 

To showcase RefactoringMiner++, we present two versions of a short sample C++ program. This example includes multiple refactorings and, additionally, a bug fix as well as a new feature. The code for the two program versions is shown in Listings \ref{listing:com1} and \ref{listing:com2}, respectively.
In particular, the \texttt{Circle} class should work as a utility class in the new version. Therefore, it was renamed accordingly and its methods have been changed to static member functions. In the second version, the bug in the \texttt{calcCircumference} method is fixed and a new member function \texttt{calcSectorArea} to get the area of a sector was added. Moreover, a few more refactorings were applied to keep naming conventions consistent throughout the code.

Table \ref{tab:exampleDetectedRefactorings} lists the detected refactorings between the first and the second version, along with the lines of code affected by each refactoring. These line numbers correspond to Listing \ref{listing:com2}. In total, five lines were refactored and all refactorings were detected without false negatives. Furthermore, one line experienced a behavior-altering modification and four lines were added in the second version. None of these lines were reported as refactorings, which means that there are no false positives either. Our extension to RefactoringMiner even reports that an added method was detected spanning lines 7-10 and that a statement was modified on line 12.

\begin{listing}
\begin{minted}[xleftmargin=20pt,linenos]{cpp} 
class Circle{
  double PI = 3.14159;
public:
  double getArea(double r){
    return PI * r * r;
  }
  double calcCircumference(double radius){
    double diameter = radius * radius; //bug
    return PI * diameter;
  }
};
\end{minted}
\caption{Code before refactorings and behavior changes.}
\label{listing:com1}
\end{listing}


\begin{listing}
\begin{minted}[xleftmargin=20pt,linenos]{cpp} 
class CircleCalculator{
  inline static const double PI = 3.14159;
public:
  static double calcArea(double radius){
    return PI * radius * radius;
  }
  static double calcSectorArea(double radius,
                               double angle){
    return angle / 360 * calcArea(radius);
  }
  static double calcCircumference(double radius){
    double diameter = radius + radius;
    return PI * diameter;
  }
};
\end{minted}
\caption{Code after refactorings and behavior changes.}
\label{listing:com2}
\end{listing}

\begin{table}[b]
  \caption{Detected refactorings in the example.}
  \label{tab:exampleDetectedRefactorings}
\begin{tabular}{lc}
  \toprule
  Detected Refactoring   & Affected Lines \\
  \midrule
  Rename Class                    & 1 \\
  Add Attribute Modifier (inline) & 2 \\
  Add Attribute Modifier (static) & 2 \\
  Add Attribute Modifier (const)  & 2 \\
  Rename Method                   & 4 \\
  Add Method Modifier (static)    & 4 \\
  Rename Parameter                & 4, 5 \\
  Add Method Modifier (static)    & 11 \\
  \bottomrule
\end{tabular}
\end{table}

\section{Evaluation}

We aim to compare RefactoringMiner++ to its Java counterpart and to this end, we generate two seeded refactoring datasets, which only differ in their language and are otherwise equivalent. Each dataset comprises pairs of programs where each pair consists of an original and a refactored version. The refactored version is identical to the original except for the application of a certain type of refactoring.

Prior research has shown which refactorings are the most common ones \cite{ecoop13refactoringStudy,tse20refactoringMiner2}. In our evaluation, we focus on the 16 refactoring types that occurred more than 100 times in the Java refactoring dataset of Tsantalis et al. \cite{tse20refactoringMiner2}. For each of these frequent refactoring types, a sample for the C++ and Java dataset is produced. This step is not performed manually. Instead, the samples are generated by an LLM, ChatGPT 4o mini\footnote{\url{https://chat.openai.com/}}, to ensure an unbiased oracle.

For each refactoring type, we use the same prompt template, which instructs the LLM to create two versions of a program in Java showcasing the refactoring type and to then convert the program pair to C++. This results in four programs per refactoring type. To ensure the generated programs are suitable for comparison, the prompt constrains ChatGPT to only use Java features that can be properly mapped to C++. Furthermore, the prompt specifies that applied changes must be behavior-preserving.
For four of the 16 sample programs, it was necessary to add prompts to reemphasize specific restrictions (e.g., to write an example focusing on only one refactoring type). In four cases, it was necessary to explain the refactoring to ChatGPT (e.g., \emph{"Pull Up Method" moves a function from a child class to its parent class.}). However, for most of the generated samples (11 of 16), explaining the refactoring was not necessary and one prompt sufficed to obtain the requested programs.


We use RefactoringMiner to find the refactorings in each Java program pair and we repeat this process for the C++ programs using our tool. The two refactoring detectors successfully located all seeded refactorings and, thus, produced equivalent results. This shows that RefactoringMiner++ can be used to detect refactorings in C++ programs---at least for simple programs.

\section{Discussion and Limitations}

While C++ and Java share many similarities, 
there are also key differences. 
This section highlights major differences as well as corresponding workarounds and current limitations of our tool.

To avoid naming conflicts, C++ has namespaces, while Java offers packages for this purpose. We translate namespaces in C++ programs to packages, so RefactoringMiner understands them. In C++, one file may comprise multiple namespaces. Java only allows one package per file, but this does not cause any issues for our tool.

C++ allows variables and functions to be declared at the namespace level outside of any class, whereas Java does not support this. We create artificial classes which have the purpose of wrapping parsed variables and functions that otherwise would not belong to any class. We map each namespace to a distinct artificial class and this approach even integrates global variables and functions seamlessly into RefactoringMiner's model structure. This way, RefactoringMiner++ can even detect refactorings outside of classes.

Compared to C++, Java also lacks the keyword \texttt{struct}. As ``[t]here is no fundamental difference between a \texttt{struct} and a \texttt{class}'' \cite{stroustrup22tourOfCpp}, we represent structs in the same way as classes in RefactoringMiner's model, while taking access specifiers---the only difference between these two keywords---into account.

Another C++ feature that is not present in Java is multiple inheritance. This feature causes problems for a Python port of RefDetect \cite{bscThesis24pythonWithRefDetect}. Being based on Java, RefactoringMiner's class representations allow up to one base class. However, the model also includes a \texttt{UMLGeneralization} list where RefactoringMiner++ stores all parent child relationships. This way, our tool accepts programs where multiple inheritance is present and it can even detect refactorings in such cases. For instance, attributes that are pulled up to different base classes, are reported as such by RefactoringMiner++.

Just as inheritance, template metaprogramming has the purpose of minimizing code duplication. Java does not offer templates, but generics instead. While these concepts are not equivalent, they share similarities \cite{acmSigplan04javaGenericsVsCppTemplates}. Therefore, RefactoringMiner++ represents templates in a C++ program as generics. Thus, RefactoringMiner++ can parse programs which contain templates. While we have seen successful refactoring detection in C++ programs with templates, further testing of this feature is required.

While developing RefactoringMiner++, we noticed that libClang does not always provide all AST nodes. In particular, nodes for some C++ keywords, such as \texttt{decltype} or \texttt{noexcept}, are missing. Therefore, we apply a workaround, in which our tool parses source code tokens to obtain all the necessary information.

The initial version of our tool has proven that RefactoringMiner can be extended to C++ programs and we expect RefactoringMiner++ to work with C++-compliant C programs as well. However, there is still room for improvement. Our tool's main limitation is that currently only two versions of a single C++ file can be compared. In most C++ programs, however, code is split across multiple header and implementation files. Support for multiple files is currently being worked on. Furthermore, lambda functions are not supported by RefactoringMiner++ at the moment. A further currently unsupported feature is the definition of classes inside of functions as well as classes nested inside a class. We plan to add these features in a future version as well.

We see further potential for improvement concerning our RefactoringMiner extension to report behavior-altering code modifications. Such code changes are currently only reported for lines which do not experience a refactoring. For lines of code, where both a refactoring and a behavior-altering change are applied, only the refactoring is reported. In future, we plan to take advantage of RefactoringMiner's fine-grained model to fix this issue.

We also acknowledge limitations in our evaluation. As other seeded refactoring datasets, our samples involve isolated refactorings. In real-life projects, however, refactorings often do not occur in isolation from behavior-altering modifications. Such tangled commits pose a challenge to refactoring detection \cite{tse20refactoringMiner2,ieeeAccess21refdetect,saner24kotlinWithRefDetect,tosem24refactoringMiner3}. Mining open-source C++ projects and manually verifying refactorings detected by our tool is an important future project. This approach would also address the second shortcoming of our dataset---its size.

To the best of our knowledge, our method to generate a refactoring dataset for evaluation is new. The initial prompts asking the LLM for a refactoring example only differed in the name of the refactoring type. As the LLM is responsible for generating samples, this approach reduces the dependency on experts who determine the ground truth in the refactoring dataset. Thus, experimenter bias is reduced. A more extensive use of the LLM would also increase our dataset's size. This way, the existence of all refactoring types of interest can be ensured, while real-world projects might lack refactorings of specific types.

Future research could also use LLMs to render refactoring detection more challenging by enriching existing refactoring datasets with more code modifications. This approach could benefit from refactoring-aware LLMs (cf. Pomian et al.~\cite{fse24llmRefactoringAssistant}) and would address the ``need [for] more challenging benchmarks''~\cite{tosem24refactoringMiner3}.

By open-sourcing our tool, we aim to support future research that depends on mining refactorings. Furthermore, our tool represents a benchmark for developers of other C++ refactoring detectors.

\section{Conclusion}

In this work, we present RefactoringMiner++, which---to the best of our knowledge---is the first publicly available refactoring detection tool for C++ programs. Our tool leverages its state-of-the-art Java counterpart, RefactoringMiner, and extends it to report behavior-altering code modifications in addition to refactorings. To obtain an evaluation dataset, we tread a new path by generating a small seeded dataset of equivalent Java and C++ refactorings using a large language model. On this dataset, our tool achieves the same results as RefactoringMiner. We acknowledge that a thorough evaluation involving real-world refactoring samples is needed and represents an important topic of future research. Future efforts will also address our tool's current limitations identified within this work.

\bibliographystyle{ACM-Reference-Format}
\bibliography{_REFERENCES}

\appendix

\section{Detailed Evaluation Results}

In this appendix, we present our evaluation results in more detail. The results shown in Table \ref{tab:evaluationWithLlm} can also be found in the readme file of our tool's GitHub repository.
The first column of Table \ref{tab:evaluationWithLlm} shows for which refactoring types the LLM was prompted to produce sample programs.
The second column indicates which refactoring types were known by the LLM without any explanation. Twelve refactoring types did not require a description.
The third column specifies for which refactoring types a single prompt sufficed to obtain sample programs. Only for four refactoring types, we had to write at least one additional prompt.
The last column shows that every sample program lead to equivalent results in RefactoringMiner and RefactoringMiner++.

\begin{table*}
  \caption{Refactoring types for which a test case was generated by ChatGPT. }
  \label{tab:evaluationWithLlm}
\begin{tabular}{lccc}
  \toprule
  \multirow{2}{*}{Detected Refactoring}    &
  No explanation of &
  Single prompt &
  Same result \\
  & refactoring needed & sufficed & in both tools \\
  \midrule
  Move Class              & -            & -            & $\checkmark$ \\
  Extract Method          & $\checkmark$ & $\checkmark$ & $\checkmark$ \\
  Change Variable Type    & $\checkmark$ & $\checkmark$ & $\checkmark$ \\
  Change Parameter Type   & $\checkmark$ & $\checkmark$ & $\checkmark$ \\
  Remame Parameter        & $\checkmark$ & $\checkmark$ & $\checkmark$ \\
  Change Return Type      & $\checkmark$ & $\checkmark$ & $\checkmark$ \\
  Rename Method           & $\checkmark$ & $\checkmark$ & $\checkmark$ \\
  Pull Up Method          & -            & -            & $\checkmark$ \\
  Move Method             & $\checkmark$ & $\checkmark$ & $\checkmark$ \\
  Rename Variable         & $\checkmark$ & -            & $\checkmark$ \\
  Move Field              & -            & $\checkmark$ & $\checkmark$ \\
  Change Field Type       & $\checkmark$ & $\checkmark$ & $\checkmark$ \\
  Extract And Move Method & -            & -            & $\checkmark$ \\
  Rename Field            & $\checkmark$ & $\checkmark$ & $\checkmark$ \\
  Pull Up Field           & $\checkmark$ & $\checkmark$ & $\checkmark$ \\
  Inline Method           & $\checkmark$ & $\checkmark$ & $\checkmark$ \\
  \bottomrule
\end{tabular}
\end{table*}

\end{document}